A. Y. Tonoyan

# Theoretical investigation of $^{41}$K states behavior under strong magnetic field and $\pi$ polarized laser field

## Abstract

*We have studied theoretically the behavior of $^{41}$K $D_1$ and $D_2$ lines for $\pi$ polarized resonant light in the presence of strong magnetic field when the total electronic angular momentum J and nuclear spin I are decoupled. We show that in the case of linear polarization and for $D_1$ line there are two transitions, so called guiding transitions (GT) that maintain their probabilities and frequency slopes. In Hyperfine Paschen-Back (HPB) regime other transitions are coming together to those GTs, making 2 groups (4 in each). Each transition in the group has the same frequency slope and probability as the GT in their group. It is demonstrated that from 12 ($D_1$) and 20 ($D_2$) initially allowed Zeeman transitions (taking into account the selection rules) at low B-field, only 8 transitions in each D line remain in absorption spectra at B > 200 G. A complete HPB regime for relatively low magnetic fields B ~ 400 G has been observed. This value is the smallest for all alkaline metals.*

**Introduction:** The influence of magnetic field on the medium and particularly on the atomic transitions is of great importance for a wide range of applications, starting from magnetometry, spectroscopy, space research, medicine, etc.[1-3]. There is a range of studies concentrated on the investigation of weak magnetic field influence on atomic transitions [4]. However, the behavior of alkali atoms in strong magnetic fields has been under attention only recently. The increasing interest is due to the fact that the external magnetic field eases the study of atomic states structure and transitions, as it separates degenerate states well from each other making possible their experimental detection. Investigations on atomic states of $^{85}$Rb, $^{87}$Rb, $^{133}$Cs and $^{39}$K are already published in [5-9, 11]. However, up to date there is not sufficient data on $^{40}$K and $^{41}$K isotopes as they are of low densities in nature (6.3% and 0.01%, respectively). In the current paper we present for the first time our theoretical studies of the $^{41}$K isotope $D_1$ and $D_2$ lines in the presence of external magnetic field while interacting with $\pi$ ($B \parallel E$) polarized laser field.

**The essence:** To understand the behavior of the alkali atoms, let us see what is happening to the quantum numbers. The splitting of atomic levels in weak magnetic fields is described by the total angular momentum $F = J + I$ of the atom and its projection $m_F$, where $J = L + S$ is the total angular momentum of electrons and $I$ is the nuclear spin. In the Hyperfine Paschen-Back (HPB) regime $J$ and $I$ become decoupled and the splitting of the atomic levels is described by their projections $m_J$ and $m_I$. For alkali metals the HPB regime takes place at fields $B \gg B_0 = A_{hfs}/\mu_B$, where $A_{hfs}$ is the ground-state hyperfine coupling coefficient (magnetic dipole constant) and $\mu_B$ is the Bohr magneton. For $^{41}$K isotope it is easy to calculate $B_0 = 90$ G. The relevant atomic configuration is presented in Figure 1 in the basis of $J$ and $I$ quantum numbers for $D_1$ and $D_2$ lines. Below we show that at strong external magnetic fields initially strong transitions weaken, instead, transitions that were negligibly weak (or even initially forbidden) become stronger and visible. The latter ones we call initially forbidden further allowed (IFFA) transitions, labeled as 3 and 11 in Figure 1. The arrows there show the transitions that remain at strong magnetic fields. Numbers in rectangles correspond to these transitions. Also, corresponding transitions in the $F, m_F$ basis are shown for $D_1$ and $D_2$ lines in Table 1 and Table 2 respectively.

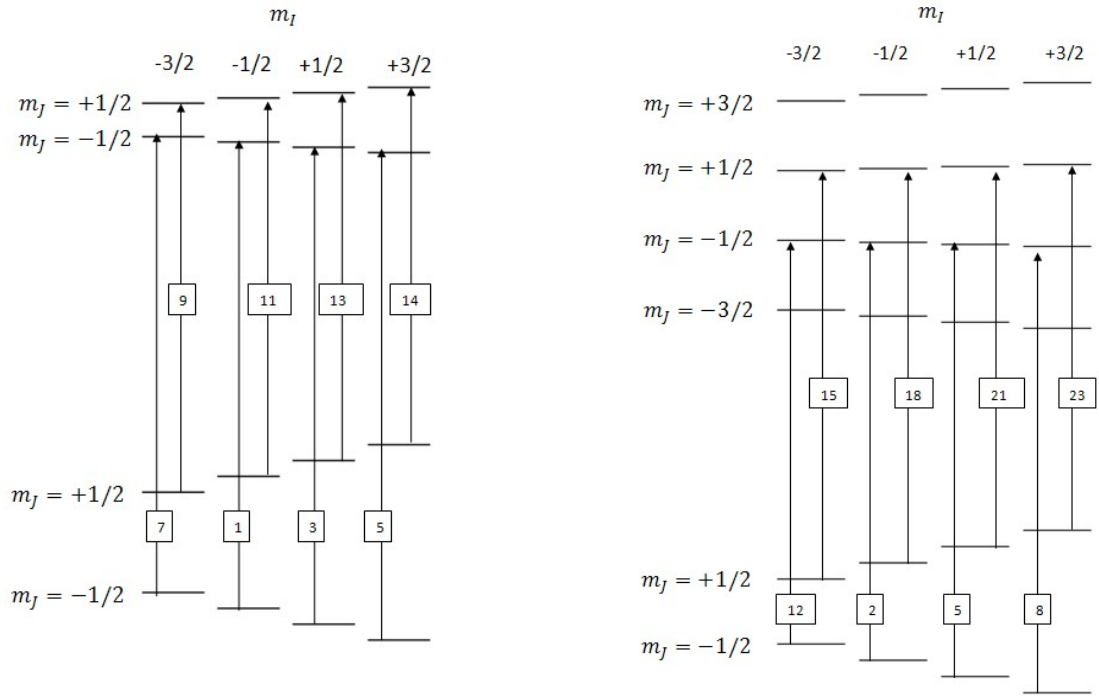

Figure 1: Relevant atomic levels and transitions in the basis of the quantum numbers $m_J$ and $m_I$. The left configuration corresponds to the $D_1$ line, the right one to the $D_2$ line. The numbers on the transitions correspond to the transitions that remain in strong magnetic fields (when $B \gg 90$ G).

There is another phenomenon that has been observed, while investigating alkali atoms $D_1$ line behavior in the $\pi$ polarized laser radiation field. There are transitions that are not changing their probabilities and maintain transition shift slopes, which we call guiding transitions (GT). The two GTs, labeled as 7 and 14 in Figure 1, maintain their probabilities and shift slopes in a very large range of external magnetic field (from 0 G to ~ 250 kG). The upper limitation is due to the fine structure constant $\alpha$.

| $|F = 1, m_F\rangle \rightarrow |F', m_{F'} = m_F\rangle$ | | | | | |
|---|---|---|---|---|---|
| $m_F \rightarrow$ | -2 | -1 | 0 | +1 | +2 |
| $F' = 2$ | | 2 | 4 | 6 | |
| $F' = 1$ | | 1 | 3 | 5 | |
| $|F = 2, m_F\rangle \rightarrow |F', m_{F'} = m_F\rangle$ | | | | | |
| $m_F \rightarrow$ | -2 | -1 | 0 | +1 | +2 |
| $F' = 2$ | 7 | 9 | 11 | 13 | 14 |
| $F' = 1$ | | 8 | 10 | 12 | |

Table 1: $D_1$ line ($|J = 1/2\rangle \rightarrow |J' = 1/2\rangle$), labels for π transitions corresponding to the labels in the rectangles in Figure 1.

| $|F = 1, m_F\rangle \rightarrow |F', m_{F'} = m_F\rangle$ | | | | | | | |
|---|---|---|---|---|---|---|---|
| $m_F \rightarrow$ | -3 | -2 | -1 | 0 | +1 | +2 | +3 |
| $F' = 3$ | | | 3 | 7 | 10 | | |
| $F' = 2$ | | | 2 | 6 | 9 | | |
| $F' = 1$ | | | 1 | 5 | 8 | | |
| $F' = 0$ | | | | 4 | | | |
| $|F = 2, m_F\rangle \rightarrow |F', m_{F'} = m_F\rangle$ | | | | | | | |
| $m_F \rightarrow$ | -3 | -2 | -1 | 0 | +1 | +2 | +3 |
| $F' = 3$ | | 12 | 15 | 19 | 22 | 24 | |
| $F' = 2$ | | 11 | 14 | 18 | 21 | 23 | |
| $F' = 1$ | | | 13 | 17 | 20 | | |
| $F' = 0$ | | | | 16 | | | |

Table 2: $D_2$ line ($|J = 1/2\rangle \rightarrow |J' = 3/2\rangle$), labels for π transitions corresponding to the labels in the rectangles in Figure 1.

**The model:** The theoretical model is based on the Hamiltonian for the D line of alkali atoms for any static magnetic field [10]. The diagonal terms of the Hamiltonian have the form

$$\langle F, m_F | H | F, m_F \rangle = E_0(F) - \mu_B g_F m_F B_z, \tag{1}$$

where $E_0(F)$ is the atomic energy in the absence of magnetic field $B_z$, $g_F$ is the corresponding Landé factor. The off-diagonal matrix elements may be non-zero only between $\Delta F = \pm 1, \Delta m_F = 0$.

$$\langle F-1, m_F | H | F, m_F \rangle = -\frac{\mu_B}{2}(g_J - g_I)B_z \left(\frac{[(J+I+1)^2 - F^2][F^2 - (J-I)^2]}{F}\right)^{1/2} \left(\frac{F^2 - m_F^2}{F(2F+1)(2F-1)}\right)^{1/2} \tag{2}$$

We use the eigenvalues to calculate the energy levels of the excited and the ground states and the state vectors are expressed in terms of the unperturbed atomic state vectors

$$|\Psi(F_e, m_e)\rangle = \sum_{F_e'} c_{F_e F_e'} |F_e', m_e\rangle \tag{3}$$

and

$$|\Psi(F_g, m_g)\rangle = \sum_{F_g'} c_{F_g F_g'} |F_g', m_g\rangle \tag{4}$$

where the sums are only over the state vectors having the same $m_F$ since the perturbation introduced by the magnetic field couples only sublevels with $\Delta m_F = 0$.

The energy shifts calculated by Eqs. (1-4) is presented in Figure 2. As we see, starting from relatively not strong magnetic fields (compared to Rb [5-7] and Cs [8, 11]), at around 300 G and 400 G for $D_1$ and $D_2$ lines, respectively, this dependence becomes linear. The mentioned difference (300 G and 400 G) is caused by the presence of electric quadrupole constant ($B_{hf}$) in $P_{3/2}$ states, which is absent for $J = 1/2$ states. Note, that the calculations show that this linearity goes up to 250 kG, however, this is not presented here in order to have a better view on the transition shifts behavior. The dashed lines in the figures represent the transitions, whose probabilities tend to zero and they are not visible in the experiments under strong magnetic fields. The solid lines are those transitions whose probabilities are not negligible and they can be seen in the experiments. As mentioned above, the definition of the numbers with the corresponding transitions in the $F, m_F$ basis is shown for $D_1$ and $D_2$ lines in Table 1 and Table 2, respectively. It is worth to mention, that in our previous experiment performed using nanocells [11], one can determine the 41K transitions in the atomic absorption spectrum [9], which, however, are much weaker compared to the $^{39}$K atomic transitions, as the density of the former is much smaller than that of $^{39}$K.

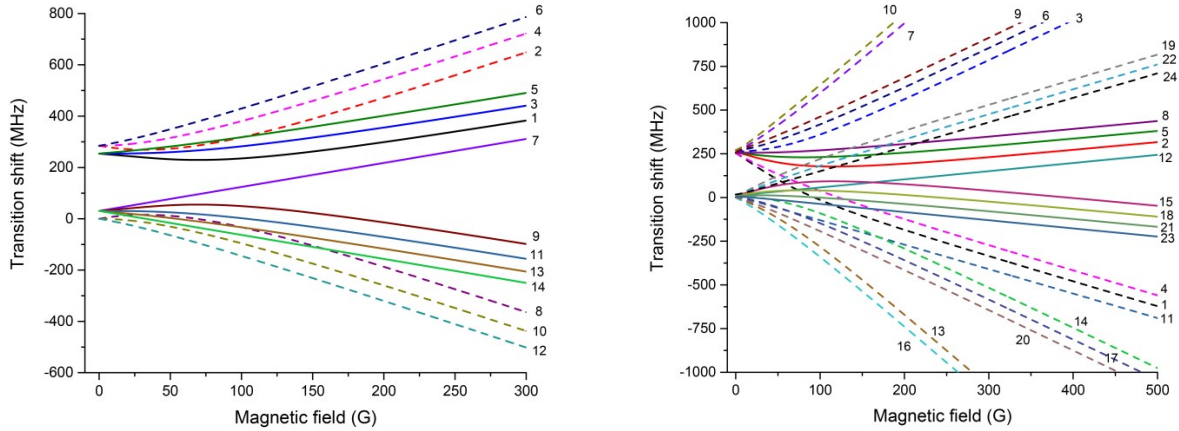

*Figure 2: Magnetic field dependence of the energy shifts of the $^{41}$K isotope $D_1$ (left) and $D_2$ (right) lines. The solid lines describe the transitions which exist in HPB regime, while dashed lines describe transitions whose probabilities tend to zero in HPB regime. The zeros are $F=2$, $m_F=-1$ to $F'=1$, $m_{F'}=-1$*

The intensities of the transitions are calculated and presented in Figure 3. As is well seen for the $D_1$ (the left figure) and $D_2$ (the right figure) lines the transitions that were strong in the presence of the weak magnetic field ($B < 80$ G for $D_1$ line and $B < 10$ G for $D_2$ line, i.e. dashed lines), become negligibly small at $B > 300$ G and $B > 400$ G, respectively, therefore they are no more visible in the experiments, while IFFA as well as less probable transitions become dominant.

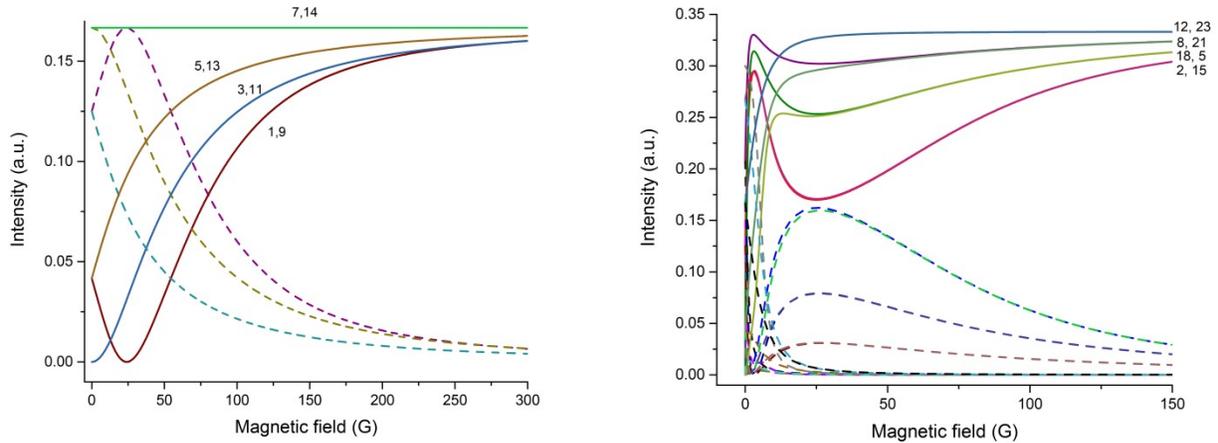

*Figure 3: Transition intensities dependence on magnetic field for $D_1$ (left) and $D_2$ (right) transitions of $^{41}$K isotope.*

The reason for having only 8 transitions for $D_1$ and 8 transitions for $D_2$ lines for $\pi$ polarized laser field is that in rather strong ($B \gg B_0$) external magnetic fields $F$ is no longer a "good" quantum number. The "good" quantum numbers become $I$ and $J$ (with their projections). GT appears because of the atomic states coupling (in an external magnetic field). Only the states which satisfy $\Delta F = \pm 1$, $\Delta m_F = 0$ selection rules are coupled. Coupling is the superposition of the quantum states. Since the quantization axis is the external magnetic field, only $\Delta m_F = 0$ states can be coupled (as was mentioned above). Not to have the same states $\Delta F$ should be $\pm 1$ and not zero (Table 3). The GTs exist exceptionally between two uncoupled states. Hence, for $^{41}$K and for $\pi$ ($B \| E$) polarization radiant field GTs are $S_{1/2}(F = 2, m_F = -2) \to P_{1/2}(F' = 2, m_{F'} = -2)$ and $S_{1/2}(F = 2, m_F = +2) \to P_{1/2}(F' = 2, m_{F'} = +2)$ transitions.

| $m_F \to$ | −2 | −1 | 0 | +1 | +2 | | $m_F \to$ | −3 | −2 | −1 | 0 | +1 | +2 | +3 |
|---|---|---|---|---|---|---|---|---|---|---|---|---|---|---|
| | \multicolumn{5}{c}{$S_{1/2}$ states} | | | \multicolumn{7}{c}{$P_{3/2}$ states} |
| $F = 2$ | UC | C | C | C | UC | | $F = 3$ | UC | C | C | C | C | C | UC |
| $F = 1$ | | C | C | C | | | $F = 2$ | | C | C | C | C | C | |
| | \multicolumn{5}{c}{$P_{1/2}$ states} | | $F = 1$ | | | C | C | C | | |
| $F = 2$ | UC | C | C | C | UC | | $F = 0$ | | | | C | | | |
| $F = 1$ | | C | C | C | | | | | | | | | | |

Table 3: Atomic states. UC-uncoupled states, C-coupled states.

As a last step, we have calculated the spectra of $^{41}$K D$_1$ and D$_2$ lines for $B = 0$ G, $B = 200$ G, $B = 400$ G and $B = 0$ G, $B = 250$ G, $B = 500$ G magnetic fields, respectively.

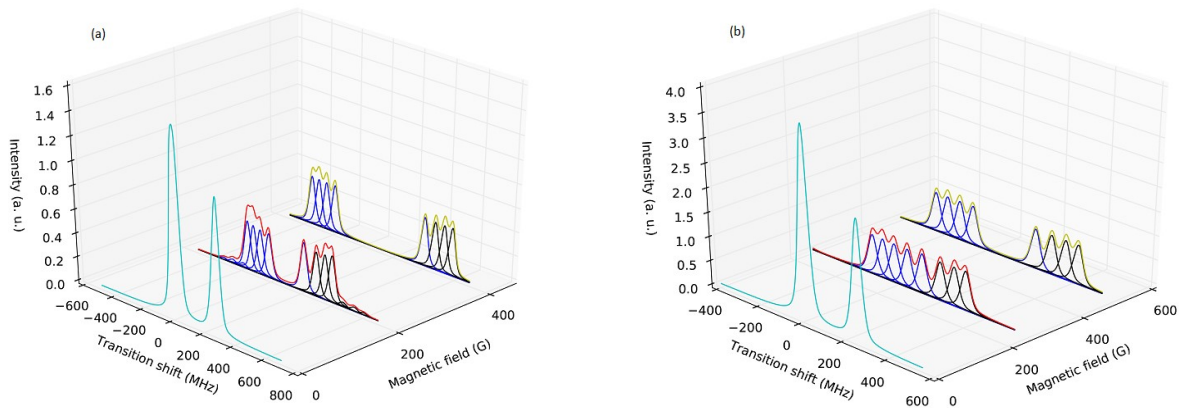

Figure 4: Spectra of $^{41}$K (a) D$_1$ line; (b) D$_2$ line for $B = 0$ G, $B = 200$ G and $B = 400$ G magnetic fields. The zeros are F=2, m$_F$=-1 to F'=1, m$_{F'}$=-1

The components coming from $F_g = 1$ are in black and the components coming from $F_g = 2$ are in blue. The components' full width at half maximum is $\omega_{FWHM} = 50$ MHz. The cumulative curves for different increasing magnetic fields are respectively in cyan, red and yellow. We claim that it is exactly what will be seen while performing experiments with the $^{41}$K isotope. Particularly, using the nano-cell with the thickness L= λ/2 (λ = 770 nm) and filled with the $^{41}$K, in the way as it was experimentally realized for the $^{39}$K atoms [9].

**Concluding**, we have studied theoretically the behavior of $^{41}$K $D_1$ and $D_2$ lines for $\pi$ polarized resonant light in the presence of strong magnetic field. We have shown that for this case in $D_1$ line there are two guiding transitions that maintain their probabilities and frequency slopes. Particularly, we explain the reason of GT existence, we observe reduction of Zeeman transitions and difference of two D lines behavior in HPB regime. Finally, we present the theoretical 3D spectra for D lines for different magnetic fields. It should be noted that a complete HPB regime for relatively low magnetic fields $B\sim$ 400 G has been demonstrated, while this value of magnetic field is the smallest for all alkaline metals.

**Acknowledgments:** The author is thankful to prof. David Sarkisyan and prof. Claude Leroy for supervising, useful and stimulating discussions. Thanks to H. Karapetyan for programming assistance and G. Hakhumyan, A. Gogyan for editing.